\documentclass[linenumbers,twocolumn]{aa}
\usepackage[utf8]{inputenc}
\usepackage{graphicx}
\usepackage{float}
\usepackage{subfigure}
\usepackage{hyperref}
\usepackage{amsmath}
\usepackage{orcidlink}
\usepackage{CJK}
\usepackage{soul}

\begin{document}
\begin{CJK*}{UTF8}{gbsn}

\titlerunning{eDIG-CHANGES III: NGC~891}
\authorrunning{L.-Y. Lu et al.}

\title{eDIG-CHANGES}
   \subtitle{III: the lagging eDIG revealed by multi-slit spectroscopy of NGC~891}

\author{
Li-Yuan~Lu~(芦李源)\inst{1,2}\orcidlink{0000-0002-3286-5346}
\and
Jiang-Tao~Li~(李江涛)\inst{2}\thanks{Corresponding author: pandataotao@gmail.com}\orcidlink{0000-0001-6239-3821}
\and
Carlos~J.~Vargas\inst{3}\orcidlink{0000-0001-7936-0831}
\and
Taotao~Fang~(方陶陶)\inst{1}\orcidlink{0000-0002-2853-3808}
\and
Robert~A.~Benjamin\inst{4}\orcidlink{0000-0002-8109-2642}
\and
Joel~N.~Bregman\inst{5}\orcidlink{0000-0001-6276-9526}
\and
Ralf-J\"{u}rgen~Dettmar\inst{6}\orcidlink{0000-0001-8206-5956}
\and
Jayanne~English\inst{7}\orcidlink{0000-0001-5310-1022}
\and
George~H.~Heald\inst{8}\orcidlink{0000-0002-2155-6054}
\and
Yan~Jiang~(姜燕)\inst{2,9}\orcidlink{0009-0003-3907-5077}
\and
Q.~Daniel~Wang\inst{10}\orcidlink{0000-0002-9279-4041}
\and
Yang~Yang~(杨阳)\inst{2}\orcidlink{0000-0001-7254-219X}
}

\institute{
Department of Astronomy, Xiamen University, 422 Siming South Road, Xiamen 361005, China
\and
Purple Mountain Observatory, Chinese Academy of Sciences, 10 Yuanhua Road, Nanjing 210023, China
\and
Department of Astronomy and Steward Observatory, University of Arizona, 933 N. Cherry Avenue, Tucson, AZ 85721, USA
\and
Department of Physics, University of Wisconsin—Whitewater, 800 West Main Street, Whitewater, WI 53190, USA
\and
Department of Astronomy, University of Michigan, 311 West Hall, 1085 S. University Ave, Ann Arbor, MI, 48109-1107, USA
\and
Ruhr University Bochum, Faculty of Physics and Astronomy, Astronomical Institute (AIRUB), 44780 Bochum, Germany
\and
Department of Physics \& Astronomy, University of Manitoba, Winnipeg, Manitoba, R3T 2N2, Canada
\and
CSIRO, Space and Astronomy, PO Box 1130, Bentley, WA 6102, Australia
\and
School of Astronomy and Space Sciences, University of Science and Technology of China, Hefei 230026, China
\and
Department of Astronomy, University of Massachusetts, Amherst, MA 01003, USA
}

\date{Received August 03, 2024; accepted October 01, 2024}

\abstract
{The kinematic information of the extraplanar diffuse ionized gas (eDIG) around galaxies provides clues to the origin of the gas.}
{The eDIG-CHANGES project studies the physical and kinematic properties of the eDIG around the CHANG-ES sample of nearby edge-on disk galaxies.}
{We use a novel multi-slit narrow-band spectroscopy technique to obtain the spatial distribution of the spectral properties of the ionized gas around NGC~891, which is often regarded as an analog of the Milky Way. We developed specific data reduction procedures for the multi-slit narrow-band spectroscopy data taken with the MDM 2.4m telescope. The data presented in this paper covers the H$\alpha$ and [\ion{N}{II}]$\lambda\lambda$6548,6583\AA\ emission lines.}
{The eDIG traced by the H$\alpha$ and [\ion{N}{II}] lines shows an obvious asymmetric morphology, being brighter in the northeastern part of the galactic disk and extending a few kpc above and below the disk. Global variations in the [\ion{N}{II}]/H$\alpha$ line ratio suggest additional heating mechanisms for the eDIG at large heights beyond photoionization. We also construct position-velocity (PV) diagrams of the eDIG based on our optical multi-slit spectroscopy data and compare them to similar PV diagrams constructed with the \ion{H}{I} data. The dynamics of the two gas phases are generally consistent with each other. Modeling the rotation curves at different heights from the galactic mid-plane suggests a vertical negative gradient in turnover radius and maximum rotation velocity, with magnitudes of approximately $3~\mathrm{kpc~kpc^{-1}}$ and $22-25~\mathrm{km~s^{-1}~kpc^{-1}}$, respectively.}
{Our measured vertical gradients of the rotation curve parameters suggest significant differential rotation of the ionized gas in the halo, or often referred to as the lagging eDIG. Systematic study of the lagging eDIG, using the multi-slit narrow-band spectroscopy technique developed in our eDIG-CHANGES project, will help us to better understand the dynamics of the ionized gas in the halo.}


\maketitle

\section{Introduction} \label{sec:Intro}


The multi-phase circum-galactic medium (CGM) is comprised of multi-phase gases, dust, cosmic ray (CR), and magnetic field in the galactic halo \citep[see the review by][]{Tumlinson17,Irwin24}.
It is often closely connected to the gas accretion and various types of galactic feedback from active galactic nucleus (AGN), massive stellar winds, and core-collapse and Type \uppercase\expandafter{\romannumeral1}a supernovae (SNe) \citep[e.g.,][]{Strickland04a, Li13b, Li15, Wang16, Li11, Li16}.

\begin{table}
\begin{center}
\caption{NGC~891: basic parameters}
\tabcolsep=7.0pt%
\begin{tabular}{lcc}
\hline
\noalign{\smallskip}
Parameter & Value & Ref. \\
\noalign{\smallskip}
\hline
\noalign{\smallskip}
R.A.(J2000) & 02$^{\rm h}$22$^{\rm m}$33$^{\rm s}$.41 & (1) \\
Decl.(J2000) & +42$^{\rm d}$20$^{\rm m}$56$^{\rm s}$.9 & (1) \\
Type & Sb & (2) \\
Distance (Mpc) & $9.1\pm 0.4$ & (3) \\
$d_{25}$ ($^{\prime}$)$^{\rm a}$ & 12.2 & (4) \\
SFR ($M_{\odot}~\mathrm{yr^{-1}}$) & 3.8 & (5)\\
$V_{\odot}$ ($\mathrm{km~s^{-1}}$)$^{\rm b}$ & 529 & (4) \\
Position Angle ($^{\circ}$) & $22$ & (1) \\
Inclination ($^{\circ}$) & $89$ & (6) \\
Total \ion{H}{I} mass ($M_{\odot}$) & $4.1\times 10^{9}$ & (6) \\
\noalign{\smallskip}
\hline
\end{tabular}\label{table:N891Properties}
\end{center}
$^{\rm a}$Observed blue diameter at the 25$^{\rm th}$~mag~arcsec$^{-2}$ isophote.\\
$^{\rm b}$Heliocentric radial velocity.\\
{\tt REFERENCES.} (1) NED; (2) HyperLeda; (3) \citealt{Radburn-Smith11}; (4) \citealt{Irwin12a}; (5) \citealt{Popescu04}; (6) \citealt{Oosterloo07}.\\
\end{table}

The $T\sim 10^{3-4}$~K warm ionized gas in the CGM is also known as the extraplanar diffuse ionized gas (eDIG).
In most of the existing narrow-band imaging observations, the eDIG is ubiquitous in actively star-forming (SF) galaxies, and often appears highly structured and distributed in a vertically extended layer above the galactic disk \citep[e.g.,][]{Hoopes96, Hoopes99, Lehnert95, Rossa00, Collins01}, with a typical exponential scale height of $\sim 1-2$~kpc \citep{Rossa03a, Jo18, Levy19, Lu23}.
The overall morphology of the eDIG is in general consistent with the galactic fountain model \citep{Shapiro76,Bregman80,deAvillez05}, while some later kinematic studies suggest that some gas in the eDIG might be infalling instead of only outflowing \citep{Heald06b,Kaufmann06}.
There are also some other multi-wavelength studies suggesting that the eDIG may co-rotate with other phases of gas, such as the neutral hydrogen and molecular gas (e.g., \citealt{Wu12,Li24}).
However, the origin and evolution of eDIG are not yet clear: How significant the accretion and/or the galactic feedback? The kinematic information of eDIG may provide the evidences.


Edge-on disk galaxies provide a direct view separating the optical line emission from the eDIG and the galactic disk.
The eDIG-CHANGES project use the novel multi-slit narrow-band spectroscopy method to study a sample of nearby edge-on disk galaxies selected from the CHANG-ES project \citep{Irwin12a,Irwin12b,Wiegert15,Irwin19,Irwin24}.
In eDIG-CHANGES~I \citep{Lu23}, we presented spatial analysis of H$\alpha$ images of the CHANG-ES galaxies obtained from the Apache Point Observatory (APO) 3.5~m Telescope \citep{Vargas19}.
Details of the project design and an example of the multi-slit narrow-band spectroscopy observation were presented in eDIG-CHANGES~II \citep{Li24}.
Compared to other methods of spatially resolved spectroscopy observations, such as using Integral Field Units (IFU) or Fabry-P\'{e}rot interferometer, our multi-slit narrow-band spectroscopy technique focuses only on the emission lines of interest, and covers a typically larger field of view (FoV; $\sim18^\prime$ for the setting used in the present paper).
The lower spatial covering fraction can be improved by using multiple offset multi-slit masks ($\sim23\%$ with the five masks; eDIG-CHANGES~II) or slightly offset the telescope pointing, and is typically sufficient to sample the radial variation of the eDIG properties.
Similar multi-slit narrow-band spectroscopy has also been adopted for spatially resolved spectroscopy studies in many previous works (e.g., \citealt{Wilson59,Crampton99,Martin04,Tran04,Arnaboldi07,Martin08,Wu12}).


In this paper, we present the detailed data reduction processes of our multi-slit narrow-band spectroscopy observations, as well as an additional case study of NGC~891 (in addition to NGC~3556 presented in eDIG-CHANGES~II).
NGC~891 is often regarded as a Milky Way analog due to their similar mass, morphology \citep[classified as Sb,][]{Nilson73, Sandage94}, and luminosity \citep{vanderKruit84}, but it has a higher star formation rate (SFR) of $3.8$~M$_{\odot}$~yr$^{-1}$ \citep{Popescu04}.
It is an ideal case for the study of the multi-phase CGM due to its high inclination angle ($i \sim 89^{\circ}$; \citealt{Oosterloo07}) and prominent multi-phase halo \citep[e.g.,][]{Strickland04a, Seon14, Boettcher16, HodgesKluck18, Das20, Reach20}.
The kinematics of the eDIG around NGC~891 have also been extensively studied using various methods \citep{Rand97, Heald06b, Kamphuis07}, which could be compared to our pilot study using the multi-slit narrow-band spectroscopy technique.
We summarize the properties of NGC~891 in Table~\ref{table:N891Properties}.

The present paper is organized as follows: In Sect.\ref{sec:MultiSlitData}, we describe the data reduction method. We then present the initial results, study the spatial variation of the ionized gas dynamics, and compare our optical data to the \ion{H}{I} data from \citet{Oosterloo07} in Sect.\ref{sec:ResultsDiscussions}. Our main results are summarized in Sect.\ref{sec:Summary}. Errors are quoted at the $1\sigma$ confidence level throughout this paper.

\section{Multi-slit spectroscopy data reduction} \label{sec:MultiSlitData}

\begin{table}
\begin{center}
\caption{Observational information}
\tabcolsep=7.0pt%
\begin{tabular}{lccc}
\hline
\noalign{\smallskip}
Date & Exposure time & Flat type$^{\ast}$ \\
\noalign{\smallskip}
\hline
\noalign{\smallskip}
24 Jan 2022 & $3\times 0.5$~h & MIS \& Sky \\
25 Jan 2022 & $3\times 0.5$~h & Sky \\
27 Jan 2022 & $2\times 0.5$~h & MIS \& Sky \\
30 Jan 2022 & $3\times 0.5$~h & MIS \& Sky \\
02 Feb 2022 & $3\times 0.5$~h & MIS \& Sky \\
\noalign{\smallskip}
\hline
\end{tabular}\label{table:ObsSummary}
\end{center}
$^{\ast}$MIS: instrument flat; Sky: twilight flat. \\
\end{table}

\begin{figure*}[ht!]
    \centering
    \includegraphics[width=0.88\textwidth]{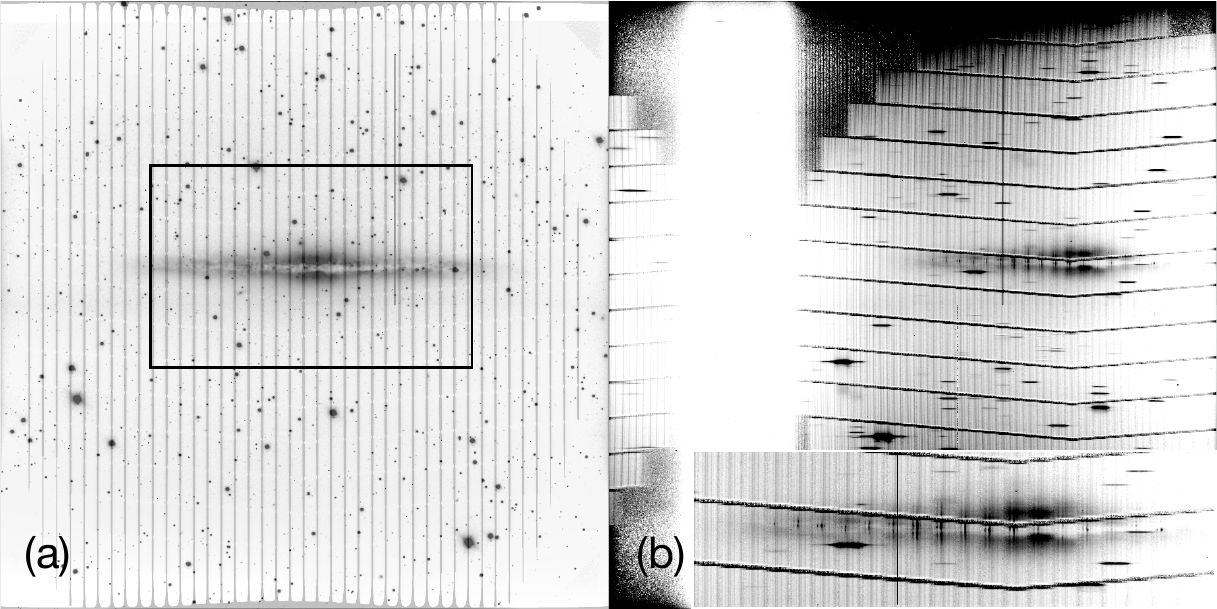}
    \caption{Raw images from the multi-slit narrow-band spectroscopy of NGC~891. The intensity scale is inverted such that bright emission is black. The instrument (OSMOS) has been rotated for $68^{\circ}$ so the galactic disk is placed perpendicular to the slits. (a) Slit image overlaid on a snapshot H$\alpha$ image, taken with the same filter KP1468 as the spectroscopy observations, and with the exposures of $3\times 60$~sec. The black box indicates the region shown in Fig.~\ref{fig:mapping}. (b) Raw spectra taken with the multi-slit mask and the KP1468 H$\alpha$ narrow-band filter. The image in the leftmost side should be higher order spectra of the rightmost slits, which are not used in the follow-up analysis. Each rectangle is a narrow-band spectrum with the $60^{\prime \prime}$ slit segment, that traces spatial length in the vertical direction and wavelength spanning the slimmer horizontal direction. The inset panel shows a zoom-in of the central region, where the three prominent emission lines (H$\alpha$, [\ion{N}{II}]$\lambda \lambda 6548, 6583$\AA) can be clearly seen as bright vertical lines.
        \label{fig:show_data}
    }
\end{figure*}

The multi-slit narrow-band spectroscopy data of NGC~891 were obtained using the Ohio State Multi-Object Spectrograph (OSMOS) imager/spectrograph \citep{Martini11} on the 2.4~m \textit{Hiltner} telescope at Michigan-Dartmouth-MIT (MDM) observatory during a series of observations.
The observation strategy and the design of the multi-slit masks are described in detail in eDIG-CHANGES~II \citep{Li24}.
Briefly, we used one of our multi-slit masks including 41 long slits uniformly distributed across the focal plane (Fig.\ref{fig:show_data}a), along with a narrow-band H$\alpha$ filter to perform imaging spectroscopy observations similar to an IFU, covering approximately $4.6\%$ of the FoV (with one mask and one telescope pointing). The narrow-band spectra typically covered the H$\alpha \lambda$6563\AA \ and [\ion{N}{II}]$\lambda\lambda$6548,6583\AA \ lines.
Details of the observations of NGC~891 are summarized in Table~\ref{table:ObsSummary}.
The instrument was rotated by $68^{\circ}$, positioning the long slits perpendicular to the disk of NGC~891 (Fig.~\ref{fig:show_data}a; position angle ${\rm PA}=22^{\circ}$ from north, measured counterclockwise).
Each single exposure takes 1800~sec with one multi-slit mask, the 4-inch KP1468 H$\alpha$ filter, the $4064\times 4064$ MDM4K CCD, and the MDM VPH red grism (resolution $R\sim 1600$).
The FoV under this setup is $\sim18.5^{\prime}$, with a plate scale of $0.273^{\prime \prime}~\mathrm{pixel^{-1}}$ at the f/7.5 focus.
The KP1468 filter has a full width at half maximum (FWHM) bandpass of 84~\AA\ centered at 6567~\AA .
The peak transmission is $\sim72\%$, and keeps $\gtrsim20\%$ in the center $\Delta\lambda\sim100~\text{\AA}$\footnote{The transmission curve of 4$^{\prime\prime}$ KP1468 is available at \href{http://mdm.kpno.noirlab.edu/filters-kpno4.html}{http://mdm.kpno.noirlab.edu/filters-kpno4.html}}.
We present an example of the raw spectra in Fig.~\ref{fig:show_data}b. Approximately 22-23 of the 41 long slits cover the optical disk of the galaxy.
Each vertical stripe represents the dispersed image of a $1.25^{\prime\prime}$-wide slit, with the spatial direction along the {\it y}-axis (perpendicular to the disk) and the wavelength direction along the {\it x}-axis.

We reduced the data partially following the OSMOS long-slit data reduction guide\footnote{\url{https://mips.as.arizona.edu/~khainline/osmos_redux.html}} developed by Kevin N. Hainline, and using a specialized pipeline developed within {\tt IRAF} \citep{Tody86,Tody93} and {\tt Python}.
Bias levels, estimated from the overscan regions of the CCD chips, were subtracted from each frame.
Since the MDM4K CCD is composed of four chips, we removed the bias from each quadrant using the {\tt colbias} procedure in {\tt IRAF} and then combined them into an image of the correct size. All the science images were also CRs removed by calculating the median of each pixel. When there are less than three images of a single setup, we used {\tt astro-SCRAPPY} \citep{McCully18} to detect and mask the CRs, which is designed to detect CRs in images based on Pieter van Dokkum's L.A.Cosmic algorithm \citep{vanDokkum01}.

The twilight flat contains some absorption lines, but are in general more uniform than the instrument flat, which has the advantage of no absorption lines. Therefore, whenever available, we do flat fielding using both the instrument and twilight flats. Field distortion was also corrected simultaneously in this process. We combined individual instrument or twilight flats using the {\tt IRAF} task {\tt flatcombine}.
When available, we used the twilight flats to estimate the full illumination pattern. In order to do this, we collapsed the flats in the spatial direction, effectively eliminating the wavelength direction.
We then divided the twilight flat trend by the instrument flat trend created in the same way, further creating a quotient that we could fit and smooth.
This quotient was multiplied by the instrument flat to correct for improper illumination.
The {\tt blkavg}, {\tt imarith}, {\tt fit1d} and {\tt blkrep} tasks were used in the above procedures.
Finally, we used {\tt response} to create the normalized response-corrected flat as the master flat frame.
For datasets containing only one type of flat frames, we simply combined and normalized the flats using {\tt flatcombine} and {\tt response}.

\begin{figure*}[ht]
    \centering
    \includegraphics[width=0.88\textwidth]{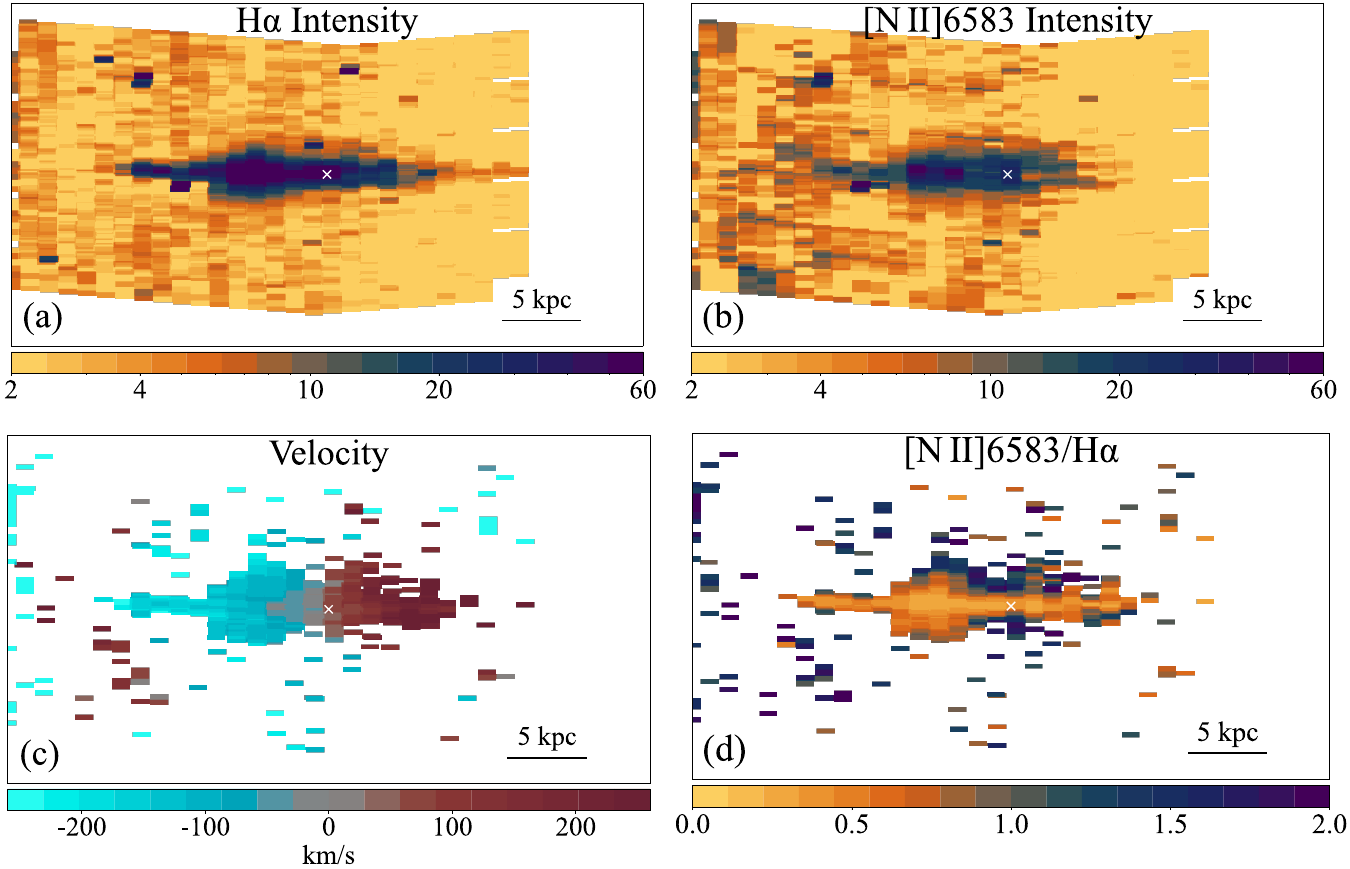}
    \caption{Calibrated data and initial results from the multi-slit spectroscopy observations of NGC 891. Panels (a) and (b) show the H$\alpha$ and [\ion{N}{II}]$\lambda$6583\AA\ line intensities in arbitrary units. Panel (c) shows the line centroid velocity jointly estimated from both the H$\alpha$ and the [\ion{N}{II}]$\lambda$6583\AA\ lines. Panel (d) shows the [\ion{N}{II}]$\lambda$6583\AA -to-H$\alpha$ line intensity ratio. The white cross in each panel marks the galactic center. The cut criteria for panels (c) and (d) are S/N$_\mathrm{H\alpha}>6$ and S/N$_\mathrm{[\ion{N}{II}]\lambda 6583\text{\AA}}>4$. The large number of spurious detections in the halo are caused by the misidentification of the residual sky lines.
        \label{fig:mapping}
    }
\end{figure*}

For wavelength calibration, we used the data of the Mercury/Neon arc lamp integrated into the Multi-Instrument System (MIS), taken with the same multi-slit mask and narrow-band filter as the science data.
By comparing the multi-slit arc spectra with the long-slit arc spectra taken with the same filter, we identified a few HgNe lines at $\lambda = 6598.95, 6532.88, 6506.53$~\AA \ at most of the slit locations (Fig.~\ref{fig:line_identification}).
For each position, we measured the central pixel of the arc lines by fitting Gaussian functions.
Since the total bandwidth of our spectra is only $\sim 100$~\AA , we transformed the pixel values to wavelengths using a linear function.
The wavelength calibration using lamp spectra in a narrow band might not be very accurate due to the small number of calibration lines and the simple linear fit.
Alternatively, night-sky emission lines provide a template for wavelength self-calibration of the science data \citep[e.g.,][]{Osterbrock96, Osterbrock97}, although they are generally less sharp than the arc-lamp lines.
The intensities of sky emission lines vary considerably with time \citep{Roach73}, on short (minutes) and long timescales, mainly due to changes in the atmosphere and in solar activity, but their wavelengths are relatively stable.
Therefore, we used the sky lines to refine and improve the wavelength calibration.
We chose several sky lines with high typical peak intensities (more than $1\times 10^{-16}~\mathrm{erg~s^{-1}~cm^{-2}~\text{\AA}^{-1}}$) from a high-resolution spectrum of optical sky emission \citep{Hanuschik03} and compared them with our data.
We identified at least three sky lines for each long slit (see an example in \ref{fig:sky_line_shift}) and measured the wavelength shift for each long slit.
The measured wavelength shifts ranged from approximately $-3$ to $+2$~\AA .
These shifts were applied in the subsequent processes.

We selected the regions above and below the central galactic disk for each long-slit to extract the background spectra.
A sixth order polynomial was used to describe the trend of background along the spatial dimension.
A lower-order polynomial may lead to a poor fit at the edges of the FoV, but it has minimal impact on the central region, where the galaxy is positioned.
The background levels were subtracted from the raw data according to the fitted trend.
We used the \texttt{Python} function, \texttt{SpectRes} \citep{Carnall17}, to resample the spectra.
The final spectra have pixel scale of $\sim 46~\mathrm{km~s^{-1}~pixel^{-1}}$ and velocity resolution of $\sim 190~\mathrm{km~s^{-1}}$.

\section{Results and discussions} \label{sec:ResultsDiscussions}

\subsection{Initial results}

\begin{figure*}[ht]
    \centering
    \includegraphics[width=0.9\textwidth]{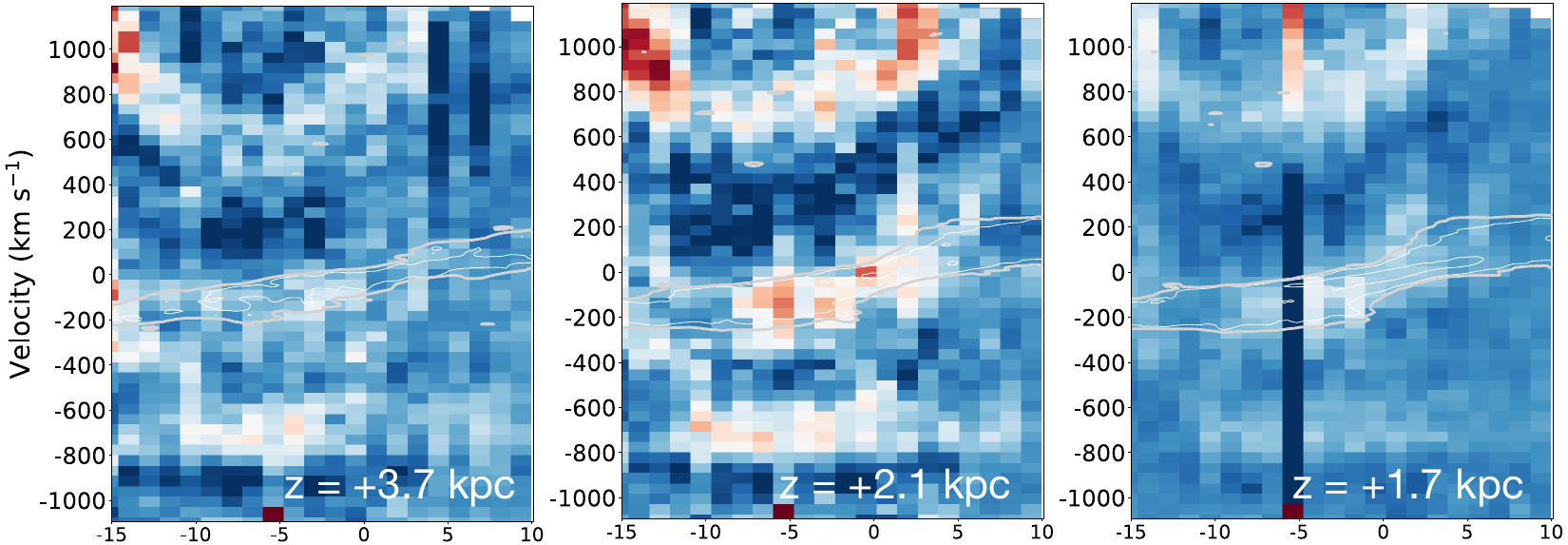}
    \includegraphics[width=0.9\textwidth]{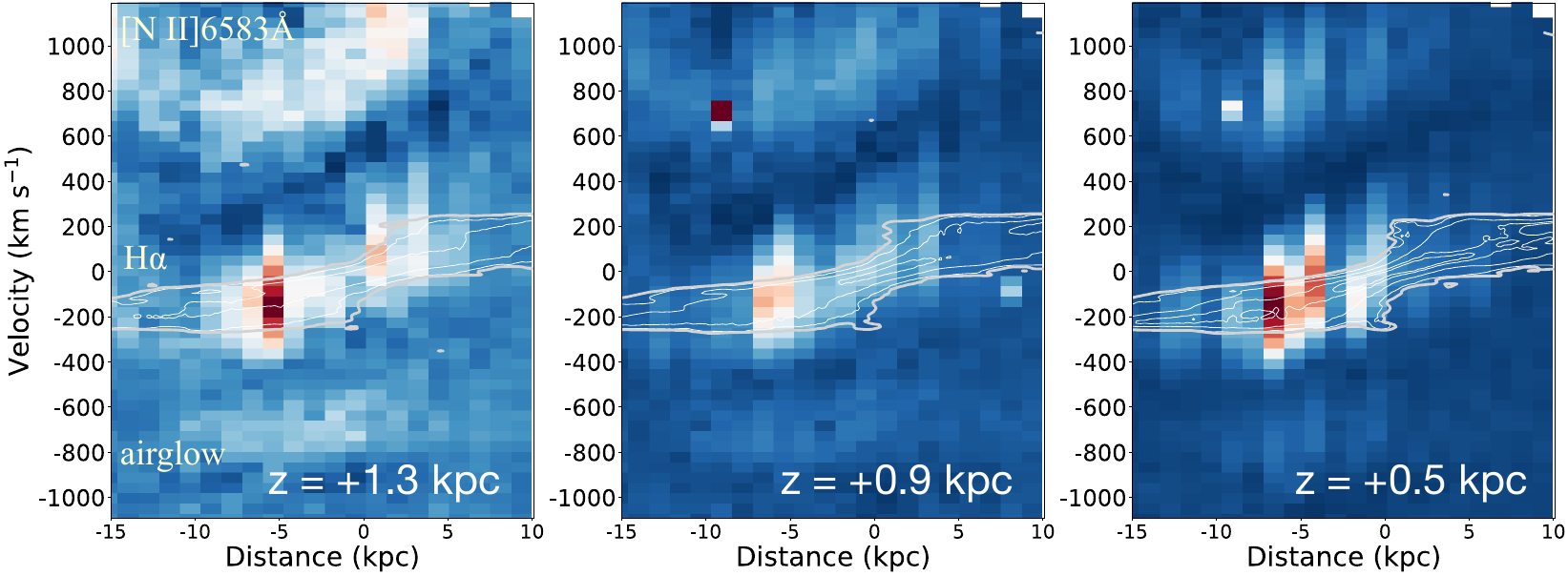}
    \caption{The position-velocity (PV) diagrams at different heights away from the disk of NGC~891 constructed with optical emission lines from our multi-slit narrow-band spectroscopy observations, with the brightness in arbitrary units. The \emph{x}-axis is the distance along the major axis. The overlaid white contours are the \ion{H}{I} 21~cm emission PV diagram constructed with high-resolution (white thin lines) and low-resolution (outermost light-grey thick line) data cubes from HALOGAS Data Release 1 \citep{Heald11,Oosterloo07}. Contour levels are $3~\sigma$ for the low-resolution data. We bin 30 pixels ($\sim 8.2^{\prime \prime}$) along the slit direction to increase the S/N. The two parallel bands of bright features are produced by the H$\alpha$ line (lower) and the [\ion{N}{II}]$\lambda 6583$\AA\ line (upper), respectively. The lowest bright band with constant velocity is contributed by the airglow, which is mixed with the [\ion{N}{II}]$\lambda 6548$\AA\ line. The pixel scales of the optical and the \ion{H}{I} spectra are $\sim 45.7~\mathrm{km~s^{-1}}$ (velocity resolution $\sim 190~\mathrm{km~s^{-1}}$) and $\sim 8.2~\mathrm{km~s^{-1}}$ (velocity resolution $\sim 16.4~\mathrm{km~s^{-1}}$), respectively, due to which the bright features on the PV diagram have different vertical extents. The slope of the two PV diagrams are comparable to each other, indicating the two gas phases have comparable rotation velocities, and no signatures of global outflows/inflows have been revealed.
    \label{fig:PVdiagram}
    }
\end{figure*}

\addtocounter{figure}{-1}
\begin{figure*}[ht]
    \centering
    \includegraphics[width=0.9\textwidth]{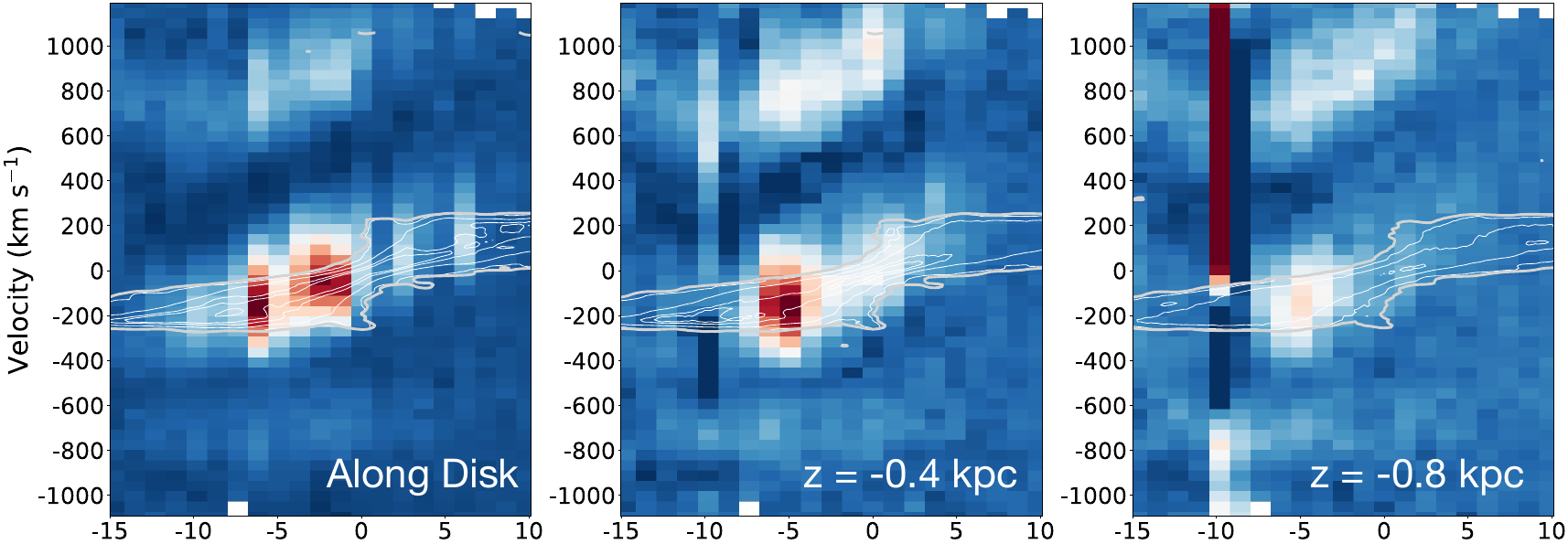}
    \includegraphics[width=0.9\textwidth]{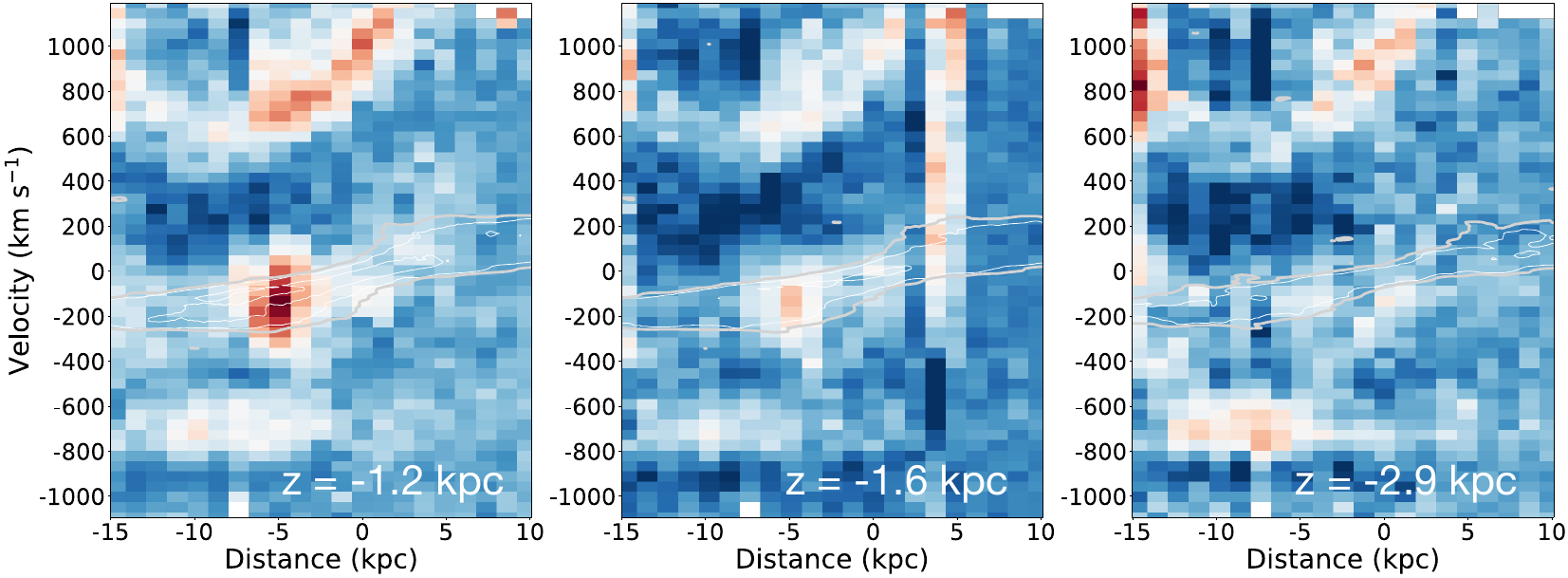}
    \caption[]{--Continued.}
    \label{fig:PVdiagram-continued}
\end{figure*}

We measured the emission-line characteristics by fitting Gaussian functions.
We fitted the H$\alpha$ and [\ion{N}{II}]$\lambda 6583$\AA\ lines simultaneously, assuming they have the same recession velocity and FWHM (in units of $\mathrm{km~s^{-1}}$).
We ignored the [\ion{N}{II}] $\lambda 6548$\AA\ line, because at most slit locations, it is too weak and always highly contaminated by a strong sky line.
We used a linear function to model the continuum in the narrow bandwidth.
The model thus includes at least six variables: amplitudes of the two Gaussian components, the centroid velocity, $\sigma$ of the Gaussian component, slope and intercept of the linear function.
We fit the above model to the spectra using the curve fitting code {\tt LMFIT}\footnote{\href{https://lmfit.github.io/lmfit-py/}{https://lmfit.github.io/lmfit-py/}} \citep{Newville14, Newville24}, with a Levenberg-Marquardt least-squares fit to minimize the $\chi ^{2}$.

The images of the H$\alpha$ and [\ion{N}{II}]$\lambda 6583$\AA \ intensity, the line centroid velocity, and the [\ion{N}{II}]/H$\alpha$ line ratio are presented in Fig.~\ref{fig:mapping} using the colour maps created by \citet{English24}. A binning of 30 pixels ($\sim 8.2^{\prime\prime}$) was adopted along the slits to increase the signal-to-noise ratio (S/N). No flux calibration had been applied so the absolute intensity has an arbitrary unit. In the following figures, the northeastern part of the galactic disk, which is the approaching side of the galaxy, is shown on the left, and the southwestern part of the galactic disk is shown on the right. The center of the galaxy is slightly shifted to the right because of the non-uniform illumination across the FoV.

As shown in Figs~\ref{fig:mapping}(a) and \ref{fig:mapping}(b), the left side of the galaxy appears significantly brighter in both H$\alpha$ and [\ion{N}{II}]. This asymmetry may be partially caused by the residual non-uniform illumination of the FoV, although we have corrected for the flat fielding. However, it may also be at least partly intrinsic, as such an asymmetry has been identified in H$\alpha$ in many other works \citep[e.g.][]{Dettmar90,Rand90,Howk97}. Furthermore, similar asymmetry in NGC~891 has also been identified in multi-wavelength, including non-thermal radio continuum emission \citep{Hummel91}, \ion{H}{I} 21-cm line emission \citep{Oosterloo07}, far-ultraviolet and near-ultraviolet emissions \citep{Seon14}, and soft X-ray emissions \citep{HodgesKluck18}. All these multi-wavelength emissions, as well as the vertical extension of them, show a global positive correlation with the SF activity (e.g., \citealt{Li13b,Wang16,Rautio22,Lu23}). This possibly indicates that the approaching (left, northeastern) side of the galactic disk may have more active star formation.

As shown in Fig~\ref{fig:mapping}(c), the galaxy shows a clear rotating disk structure on the line centroid velocity map. There is also an ``X-shaped'' structure in the central area of the galaxy, probably associated with the nuclear superwind or global galactic outflow (e.g., \citealt{Bregman13,HodgesKluck18}).

Notwithstanding the absence of flux calibration and dust correction, we see a clear global spatial variation in [\ion{N}{II}] intensity relative to H$\alpha$ (Fig.~\ref{fig:mapping}d).
Since these two lines are very close in wavelength, dust correction will not significantly affect the observed line ratios.
The [\ion{N}{II}]/H$\alpha$ value increases with increasing distance from the stellar disk.
A similar pattern of the [\ion{N}{II}]/H$\alpha$ line ratio in the eDIG component of NGC~891 was also revealed in long-slit spectroscopy \citep{Rand97,Rand98} or IFU studies \citep{Boettcher16}.
In our Milky Way Galaxy, the large increases in the forbidden lines to H$\alpha$ line ratios with the vertical distance from the mid-plane are also observed \citep{Haffner99}.
As the variations in [\ion{N}{II}]/H$\alpha$ essentially trace variations in temperature \citep{Haffner09}, following
\begin{equation}
\frac{I_\mathrm{[\ion{N}{II}]6583}}{I_\mathrm{H\alpha}}=12.2T_4^{0.426}e^{-2.18/T_4},
\end{equation}
where $T_4$ is the electron temperature in units of $10^4~{\rm K}$, the pattern of [\ion{N}{II}]/H$\alpha$ here illustrates an increase in temperature with increasing height.
Furthermore, the [\ion{N}{II}]/H$\alpha$ line ratio strongly depends on the ionization parameter \citep{Kewley19}. In current photoionization models, the electron temperature changes with distance from the ionizing source.
The increasing [\ion{N}{II}]/H$\alpha$ line ratio suggests that additional mechanisms, such as the dissipation of interstellar plasma turbulence \citep[e.g.,][]{Reynolds99}, may contribute to heating the eDIG at greater heights, beyond the effects of photoionization alone.

\subsection{Lagging halo in eDIG}

\begin{figure*}[ht]
    \centering
    \includegraphics[width=1.0\textwidth]{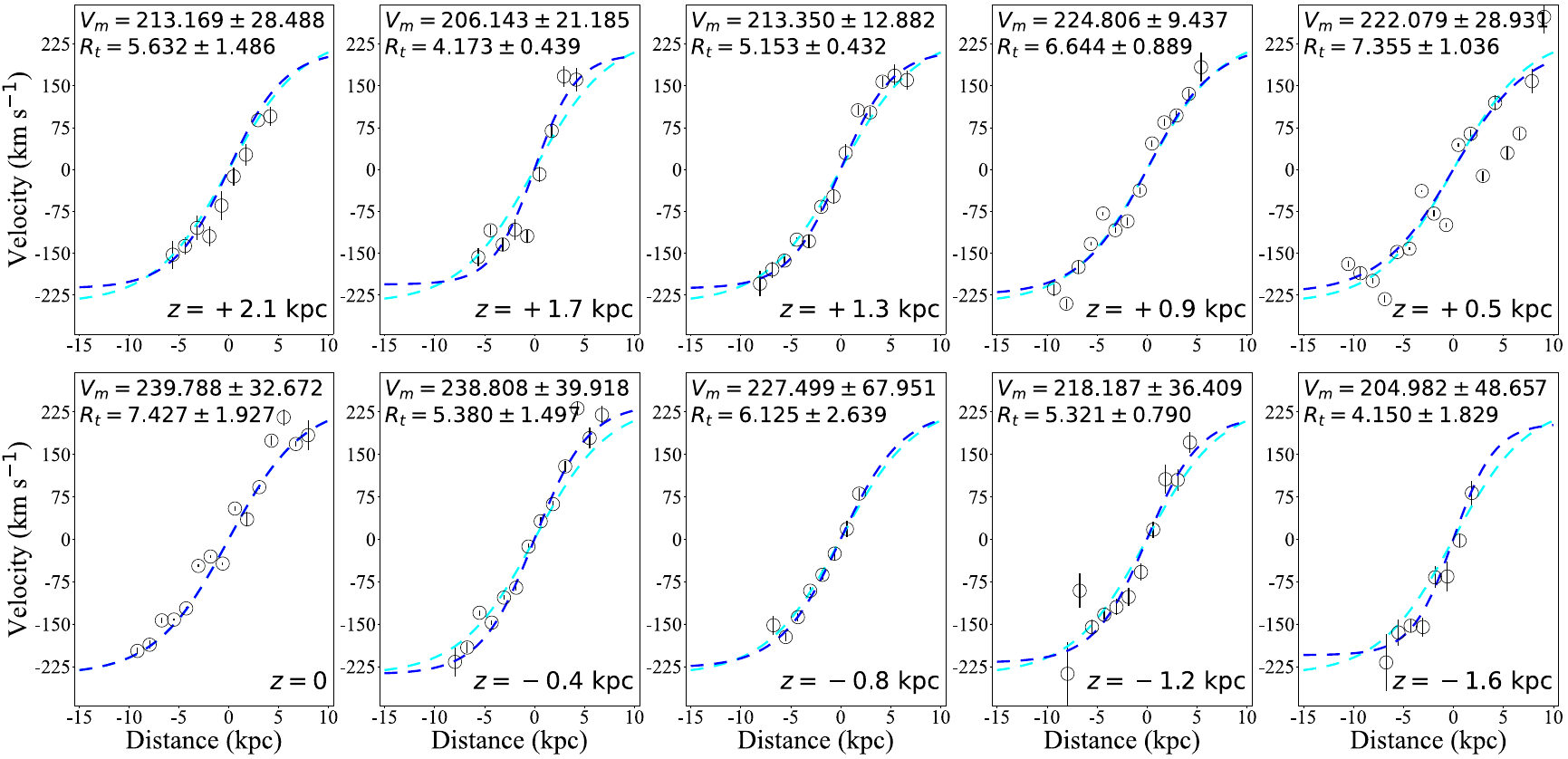}
    \caption{The rotation curves at different heights away from the mid-plane of NGC~891. The \emph{x}-axis is the distance along the major axis. The best-fit results are depicted as the blue dashed curves, while the cyan dashed curves represent the best-fit curve along the mid-plane ($z=0$). The negative values of $z$ indicate the direction below the disk or the southern half of the halo, and vice versa for positive values. 
        \label{fig:RC_fitting}
    }
\end{figure*}

We compare the kinematics of the eDIG at different heights above the galactic plane using the multi-slit narrow-band spectroscopy data. The results are presented as position-velocity (PV) diagrams in Fig.~\ref{fig:PVdiagram} and rotation curves in Fig.~\ref{fig:RC_fitting}.
In addition to our optical data, we also extract similar PV diagrams from the \ion{H}{I} 21~cm line data obtained from the Westerbork Synthesis Radio Telescope (WSRT) Hydrogen Accretion in LOcal GAlaxieS (HALOGAS) Survey (\citealt{Heald11}; the data of NGC~891 is first published in \citealt{Oosterloo07}).
The \ion{H}{I} 21~cm line PV diagram is plotted as contours on the related figure panels for comparison.
The velocity resolution of the \ion{H}{I} spectra is $\sim 16.4~\mathrm{km~s^{-1}}$, much finer than that of our optical spectra.

The panel denoted ``Along Disk'' in Fig.~\ref{fig:PVdiagram} shows the PV diagram along the mid-plane of NGC~891, with the corresponding rotation curve presented in the bottom left panel of Fig.~\ref{fig:RC_fitting} ($z=0$).
The shape of the rotation curve is typical for spiral galaxies (e.g., \citealt{Salucci07}), and shows signatures of flattening at $r\gtrsim6\rm~kpc$. However, the depth of our observations is insufficient to detect the H$\alpha$ line at large enough galactocentric radii, so the flattened part of the rotation curve is not well constrained. Nevertheless, it is clear that the maximum rotation velocity at most of the vertical distances should be $\gtrsim200~\mathrm{km~s^{-1}}$. As shown in Fig.~\ref{fig:PVdiagram}, the PV diagram of \ion{H}{I}, although with a much higher velocity resolution so in general appearing narrower, has roughly comparable slopes as the H$\alpha$ data whenever the latter is clearly detected. This apparently indicates that the two gas phases are in general co-rotating. The \ion{H}{I} data shows some elevated velocity components close to the center of the galaxy (in both radial and vertical distances), which are also shown in $^{12}$CO $J=1-0$ PV diagram \citep{Yim11}, probably suggesting the contribution from galactic nuclear outflows. This component, however, is not revealed by our H$\alpha$ data, probably because of the low velocity resolution. Such an outflow, however, is suggested by multi-wavelength observations of NGC~891 (e.g., \citealt{Bregman13,HodgesKluck18}) or other similar galaxies with disk-wide SF (e.g., \citealt{Strickland04a,Li08,Heald22}).

We measure the centroid velocities by fitting Gaussian functions on the PV diagrams using the curve fitting code {\tt LMFIT}.
The best-fit values along with their $1\sigma$ errors are plotted in Fig.~\ref{fig:RC_fitting}.
Following \citet{Yoon21}, we use a hyperbolic tangent function to fit the rotation curve at different heights from the galactic plane:
\begin{equation}
    V(R)=V_{m}\tanh{(\frac{R}{R_{t}})}, 
\end{equation}
where $R_{t}$ is the characteristic turnover radius beyond which the rotation curve becomes flat.
Different from \citet{Yoon21}, since we have little constraint on the shape of the rotation curve at large radii, and a single hyperbolic tangent function is always good enough to describe the data, we do not include an additional linear component in the fitting function.
We also applied an inclination correction to the maximum rotation velocity $V_{m}$, adopting an inclination angle of $i\sim 89^{\circ}$ (Table~\ref{table:N891Properties}). When fitting the rotation curves, we estimate the errors of the parameters by applying the same analyses to 1000 bootstrap-with-replacement sampled data. Both the measurement uncertainties and the systematic dispersion of the sample are taken into account, following a similar approach as detailed in \citet{Li13b} and \citet{Li16}.
The best-fit value of $V_m$ along the galactic mid-plane after inclination correction is $240\pm 33~\mathrm{km~s^{-1}}$, which is consistent with the measurement from the \ion{H}{I} line, $\sim 236~\mathrm{km~s^{-1}}$, as reported by \citet{Oosterloo07}.
Using the model described in \citet{Salucci07}, we obtained a virial mass of $M_{\rm vir}\sim 10^{12.4}~\rm{M_{\odot}}$ from this rotation velocity.

\begin{figure}[t!]
    \centering
    \includegraphics[width=0.47\textwidth]{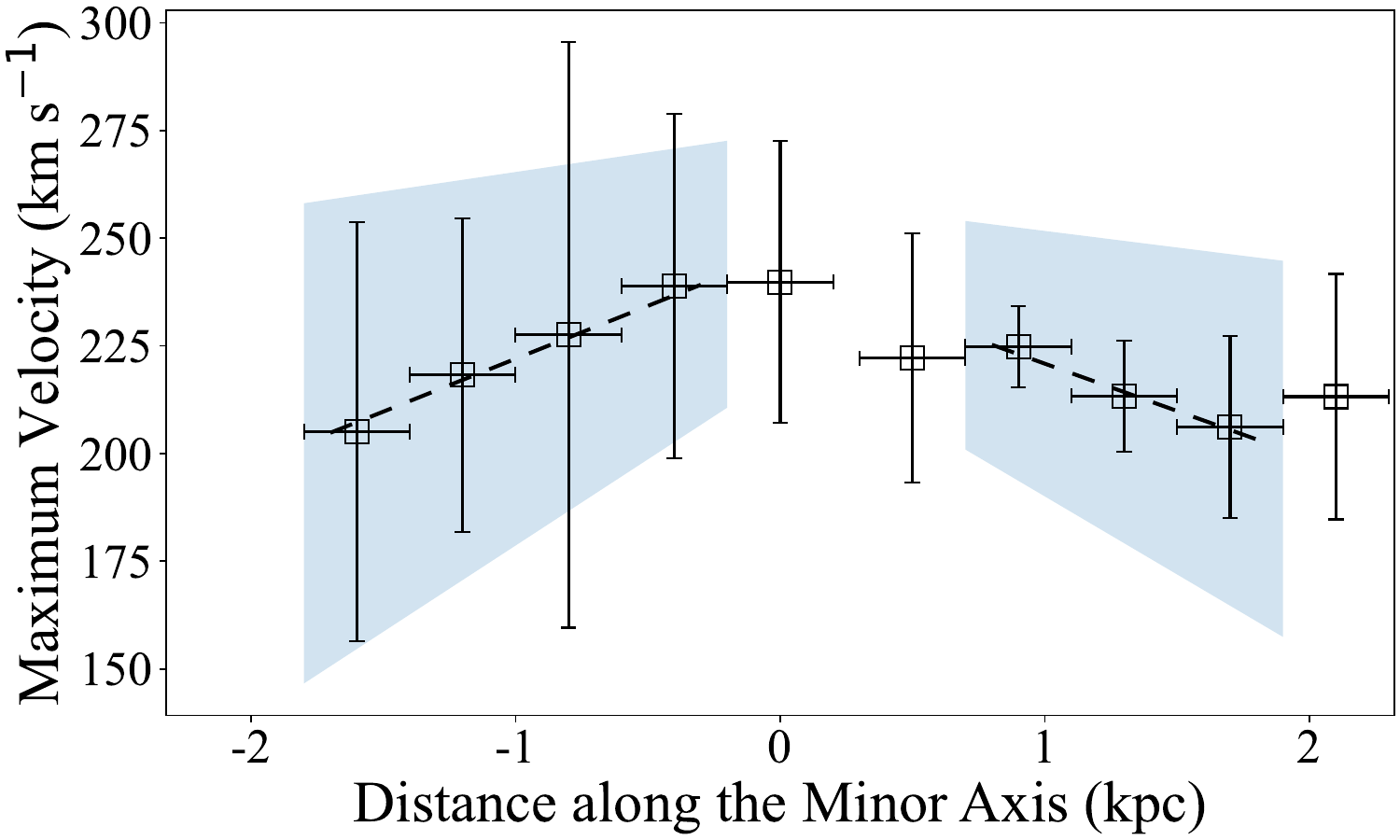}
    \includegraphics[width=0.47\textwidth]{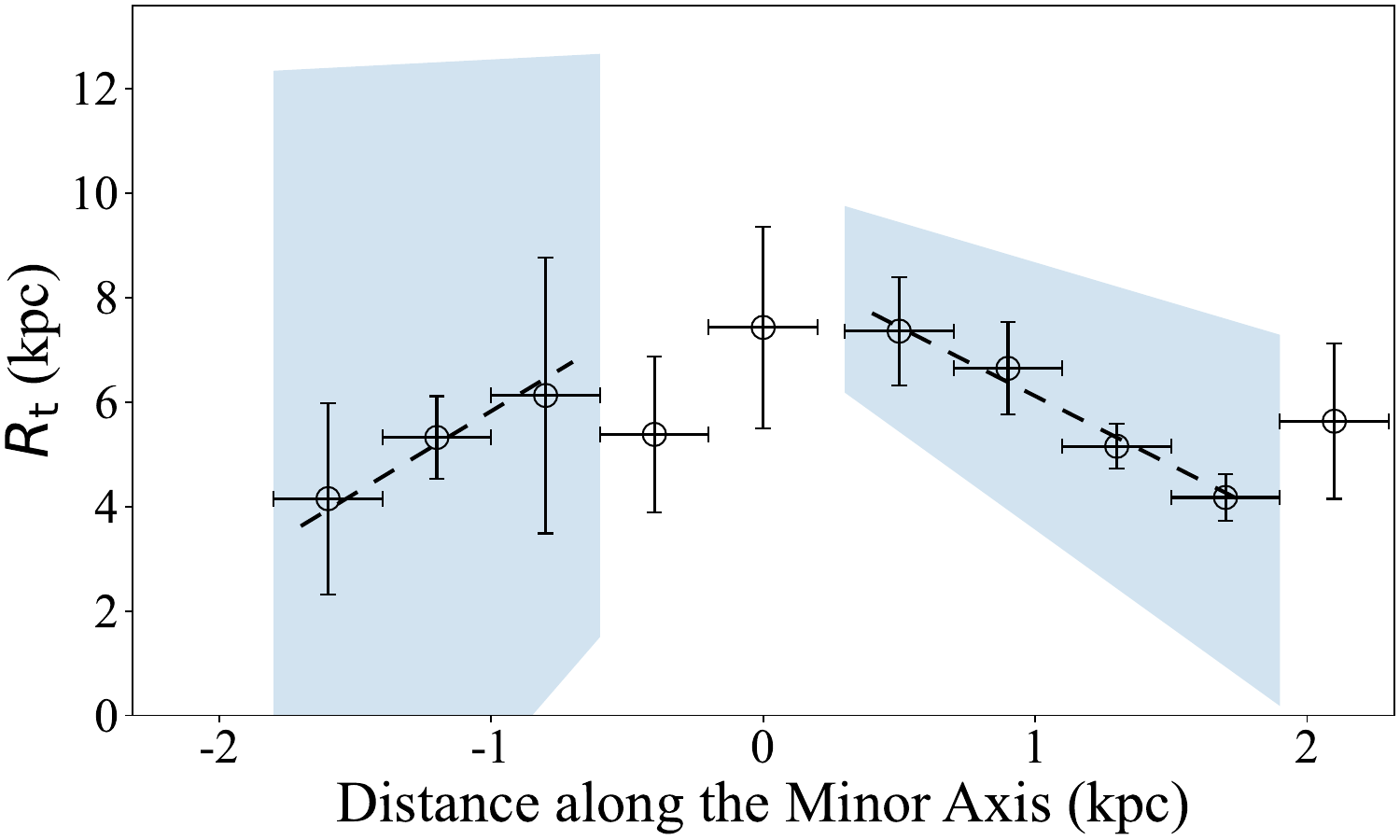}
    \caption{The maximum rotational velocities (\textit{top}; black squares) and the turnover radii (\textit{bottom}; black circles) calculated from the rotation curves at each height. The values of the velocities are inclination-corrected, adopting an inclination angle of $i=89^{\circ}$. Horizontal errors indicate the range of heights (30 pixels, $\sim 8.2^{\prime \prime}\approx 0.4$ kpc) binned along the slit direction. A bootstrap-with-replacement analysis is applied, and the best-fit results are shown as black dashed lines. The shaded blue regions mark the $1\sigma$ errors.
    The best-fit solutions show decreases in $dV_{m}/dz$ of $\sim -24.5\pm 15.5~\mathrm{km~s^{-1}~kpc^{-1}}$ (below the disk or the southern half of the halo) and $\sim -22.0\pm 14.3~\mathrm{km~s^{-1}~kpc^{-1}}$ (above the disk or the northern half of the halo), and decreases in $dR_{t}/dz$ of $\sim -3.1\pm 2.9~\mathrm{kpc~kpc^{-1}}$ (below the disk) and $\sim -2.6\pm 1.1~\mathrm{kpc~kpc^{-1}}$ (above the disk).
        \label{fig:lagging}
    }
\end{figure}

We further examine the vertical variation of the maximum rotation velocity $V_{m}$ and the turnover radius $R_{t}$ in Fig.~\ref{fig:lagging}. The vertical gradient of these parameters help us to constrain the ``lagging'' of the eDIG, which is a decrease in rotational velocity with increasing height (e.g., \citealt{Heald07}). 
As the lower halo does not show a monotonic variation of $V_{m}$ and $R_{t}$, we determine the gradient of them above a certain height from the galactic midplane. The measured vertical gradients of the turnover radius $R_{t}$ are $\Delta R_{t}/\Delta z= -3.1\pm 2.9$ and $-2.6\pm 1.1~\mathrm{kpc~kpc^{-1}}$ on the southern ($-1.6<z<-0.8$~kpc) and northern ($0.5<z<1.7$~kpc) half of the halo, respectively.
The measured vertical gradients of the maximum rotation velocity $V_{m}$ are $\Delta V_{m}/\Delta z= -24.5\pm 15.5$ and $-22.0\pm 14.3~\mathrm{km~s^{-1}~kpc^{-1}}$ at $-1.6<z<-0.4$~kpc and $0.9<z<1.7$~kpc, respectively.

Similar rotation velocity gradients of eDIG and other gas phases of NGC~891 have also been reported in many other works.
For example, \citet{Oosterloo07} determined the rotation velocity in their \ion{H}{I} observations using the envelope-tracing method \citep{Sancisi79}, and found a vertical gradient of the rotation velocity of $\Delta v_{\rm rot}/\Delta z\sim-15~\mathrm{km~s^{-1}~kpc^{-1}}$ for the halo component.
\citet{Kamphuis07} found a gradient of $-18.8\pm 6.3~\mathrm{km~s^{-1}~kpc^{-1}}$ from their imaging Fabry-P\'{e}rot spectrograph observations of the H$\alpha$ line. Furthermore, the velocity gradient has also been measured in small regions around NGC~891, although such a method may be biased due to the lack of sampling along the radial direction. For example, \citet{Rand97} found a change in velocity centroids by about $30~\mathrm{km~s^{-1}}$ from $z=1$~kpc to $z=4.5$~kpc from their long-slit spectral observations with the long-slit positioned at $R=-5$~kpc.
In an IFU observation of the northeast quadrant of the halo ($-7<R<-4$~kpc and $-4.8<z<-1.2$~kpc), \citet{Heald06b} obtained a vertical velocity gradient of $-17.5\pm 5.9~\mathrm{km~s^{-1}~kpc^{-1}}$.
Considering the large uncertainties and sometimes the different methods to characterize the rotation curve (e.g., the envelope-tracing method; the intensity-weighted-velocity method; \citealt{Sofue01}), our measured H$\alpha$ rotation velocity gradient is in general consistent with these works.

The lagging eDIG has also been revealed in many other galaxies, from either case studies of individual galaxies \citep[e.g.][]{Heald06a,Heald07}, or galaxy samples \citep[e.g.][]{Zschaechner15,Levy18,Levy19,Bizyaev17,Bizyaev22}.
For example, \citet{Levy19} studied the vertical gradients of the rotation velocity of the ionized gas in 25 edge-on galaxies selected from the CALIFA \citep{Sanchez16a} survey.
They found the distribution of $\Delta V/\Delta z$ peaked at around $-20~\mathrm{km~s^{-1}~kpc^{-1}}$, with a median value of $\langle \Delta V/\Delta z\rangle = -19^{+17}_{-26}~\mathrm{km~s^{-1}~kpc^{-1}}$.
Interestingly, all of the four galaxies having outflows in their sample are also located around this value \citep[Fig.~6a in][]{Levy19}, indicating that global outflows may not significantly affect the gas accretion.
We therefore believe that our measured rotation velocity gradient, although with low velocity resolution spectra and possibly affected by unresolved velocity distortions from the galactic outflow, is still characteristic of the gas accretion in a typical $L^\star$ galaxy.

The origin of the lags in the eDIG is complex.
Such a differential rotation is potentially linked to gravitational effects instead of SF processes.
This is confirmed in many previous works, showing no \citep{Levy19,Bizyaev22} or sometimes even negative correlation \citep{Heald07} between the lags and the the SF activity, while a positive correlation between the magnitude of the lags and both the stellar mass and central velocity dispersion (e.g., \citealt{Bizyaev22}).
\citet{Levy19} also suggested that the lags are generally consistent with the behavior induced by a thick-disk potential.
Moreover, some other effects may further affect the magnitude of the lags, such as the influence of AGN \citep{Bizyaev22} and magnetic field \citep{Jalocha12,Henriksen16}.
Quantitatively modeling of the lagging eDIG, as well as systematic comparison of the eDIG lagging model parameters to other galaxy properties or similar lag parameters of other gas phases, will be crucial to understand the shape of the gravitational potential around galaxies, as well as the gas dynamics and feedback processes.
These studies will be presented in follow-up papers of the eDIG-CHANGES project.

\section{Summary} \label{sec:Summary}

In this paper, we present our multi-slit narrow-band spectroscopy observations of the edge-on MW-like galaxy NGC~891.
We present detailed data reduction processes and images of the H$\alpha$ and [\ion{N}{II}]$\lambda 6583$\AA \ intensities (without flux calibration), the line centroid velocity, and the [\ion{N}{II}]/H$\alpha$ intensity ratio.
We observe an asymmetry in H$\alpha$ and [\ion{N}{II}] intensities and a significant vertical variation in the [\ion{N}{II}]/H$\alpha$ line ratio.
Near the galactic mid-plane, the [\ion{N}{II}]/H$\alpha$ ratio is low, consistent with the characteristic \ion{H}{II} region emission, while at larger heights ($z>1$~kpc), the ratio exceeds $\sim 1$, consistent with those expected from diffuse emission, which may suggest additional heating mechanisms for the eDIG beyond photoionization.

We construct PV diagrams using spectra from individual slits, binned with a $\sim 8.2^{\prime \prime}$ segment along the slits.
This PV diagram of the eDIG is further compared to the PV diagram of the \ion{H}{I} data constructed using the WSRT data.
The dynamics of the two gas phases are generally consistent with each other, with a maximum rotation velocity along the mid-plane of $\sim 240~\mathrm{km~s^{-1}}$.
We estimate the maximum rotation velocities at different heights above and below the disk, revealing a negative gradient of the maximum rotation velocity in the vertical direction, with a magnitude of $\Delta V_m/\Delta z\sim 22-25~\mathrm{km~s^{-1}~kpc^{-1}}$.
We also find a negative gradient in turnover radius with a magnitude of $\Delta R_t/\Delta z\sim 3~\mathrm{kpc~kpc^{-1}}$.
These indicate a significant lagging rotation of the eDIG in the halo of NGC~891, as also suggested in many other galaxies.
The lags of the eDIG should be primarily influenced by the gravitational potential, while also affected by other processes, such as the AGN feedback and galactic scale magnetic fields.
A systematic study of the lagging eDIG around nearby galaxies, using techniques like our multi-slit narrow-band spectroscopy, will help us to better understand the dynamics of the ionized gas in galactic halos.

\begin{acknowledgements}
The authors acknowledge the late R. Mark Wagner from The Ohio State University, who helped us make the multi-slit masks.
We also acknowledge Eric Galayda and Justin Rupert at the MDM observatory for the help in the multi-slit spectroscopy observations, and Prof. Filippo Fraternali from University of Groningen for providing the WSRT \ion{H}{I} data.
This work is supported by the National Natural Science Foundation of China under Nos. 11890692, 12133008, 12221003.
We acknowledge the science research grant from the China Manned Space Project with No. CMS-CSST-2021-A04.
J.T.L acknowledges the financial support from the National Science Foundation of China (NSFC) through the grant 12273111, and the science research grants from the China Manned Space Project.
R.J.D. acknowledges support from Deutsche Forschungsgemeinschaft through SFB\,1491.
Y.Y. acknowledges support from the NSFC through grant No. 12203098.
This research made use of \texttt{IRAF}, a general purpose software system for the reduction and analysis of astronomical data \citep{Tody86, Tody93}.
This work made use of \texttt{Astropy}:\footnote{http://www.astropy.org} a community-developed core Python package and an ecosystem of tools and resources for astronomy \citep{astropy:2013, astropy:2018, astropy:2022}.
This research made use of \texttt{ccdproc}, an \texttt{Astropy} package for image reduction \citep{Craig17}.
This research made use of data from WSRT HALOGAS-DR1. The Westerbork Synthesis Radio Telescope is operated by ASTRON (Netherlands Institute for Radio Astronomy) with support from the Netherlands Foundation for Scientific Research NWO.
\end{acknowledgements}

\bibliographystyle{aa}
\bibliography{main}{}

\appendix
\section{Wavelength calibration}

The process of wavelength calibration for observations with a single long-slit is well developed.
For OSMOS and the same VPH red grism on the MDM/2.4~m \textit{Hiltner} telescope, the Mercury/Neon (HgNe) or Argon/Xenon (ArXe) arc lamps integrated into the MIS are often used. For a broad band spectrum, a few tens of arc lines could typically be identified and a third order Legendre polynomial is often used to transform the pixel values to wavelengths.

\begin{figure*}[htbp]
    \centering
    \includegraphics[width=1.0\textwidth]{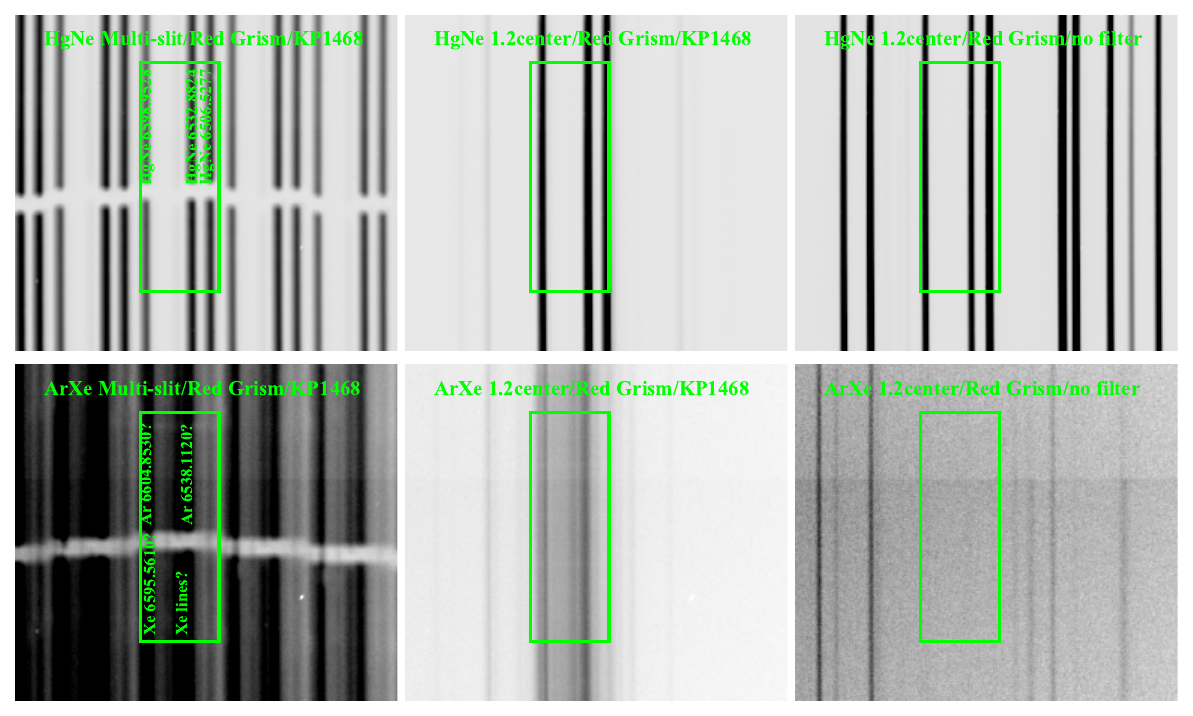}
    \caption{The identification of the HgNe (\textit{top}) and ArXe (\textit{bottom}) lines. The raw lamp data taken with (\textit{left}) the multi-slit mask and KP1468 narrow-band filter; (\textit{middle}) the long slit positioned at center and KP1468 narrow-band filter; (\textit{right}) the long slit positioned at center without filter are shown. The lime boxes mark the same positions on the CCD chip. For HgNe lines, three lines are identified, with their wavelengths noted on the \textit{top-left} panel. There are no lines strong enough being covered by the KP1468 narrow-band filter for ArXe lines.
        \label{fig:line_identification}
    }
\end{figure*}

Different from the calibration of a broad-band spectrum, wavelength calibration in a narrow band is often challenging, with two major difficulties.
The first one is identifying the arc lines, which is often rare in a specific narrow-band.
We take the lamp spectra of the HgNe or ArXe lamp using a single long slit positioned at the center of the FoV with or without the narrow-band filter KP1468, which has been used in the scientific observations.
The presence of the filter causes a slight shift of the arc lines, but we can still identify three obvious HgNe lines covered by the filter by comparing the lamp spectra with and without the filter.
The wavelengths of these three HgNe lines are $\lambda = 6598.95, 6532.88, 6506.53$~\text{\AA} (Fig.~\ref{fig:line_identification}).
We then take the multi-slit lamp spectra with the same narrow-band filter.
There is also a long slit positioned at the center of the FoV among the 41 long slits on our multi-slit mask, which produces similar arc spectra as the single long-slit.
By comparing the positions of the arc lines produced by different slits on the CCD chip, we can easily identify these three HgNe lines for each slit, which are then used for wavelength calibration.
The ArXe lamp does not produce any strong lines covered by the KP1468 filter. We can identify a few weak lines (Fig.~\ref{fig:line_identification}), but do not involve them in the process of wavelength calibration.

\begin{figure*}[htbp]
    \centering
    \includegraphics[width=0.75\textwidth]{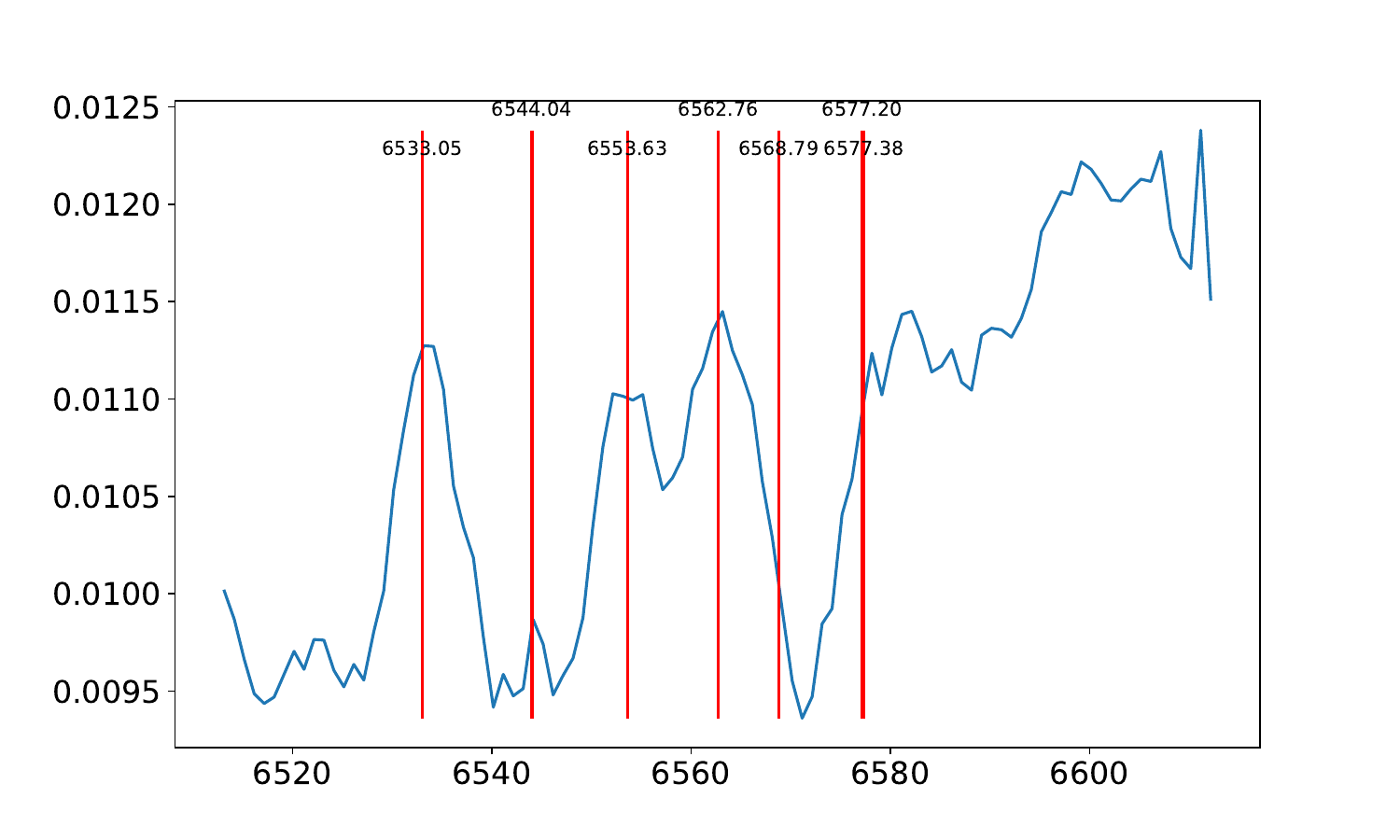}
    \caption{An example of wavelength transformation correction. The red vertical lines mark the strong sky lines with high typical peak intensities from an atlas of optical sky emission (see details in the text). The x-axis represents the wavelength, while the y-axis represents the intensity in arbitrary units. In this example, three sky lines are identified ($\lambda = 6533.05, 6553.63, 6562.76$~\text{\AA}). After correcting the wavelength transformation, the observed sky lines align correctly with the wavelengths from the atlas.
        \label{fig:sky_line_shift}
    }
\end{figure*}

The second challenge of wavelength calibration in a narrow band is wavelength transformation.
As discussed in Sect.\ref{sec:MultiSlitData}, we used a linear function for the transformation because the total bandwidth of our spectra is only $\sim 100~\text{\AA}$.
However, the small number of arc lines and the simple linear fit could result in biases in wavelength transformation, which may vary from slit to slit.
We use night-sky emission lines to further correct for these biases.
There are seven sky emission lines with typical peak intensities greater than $1\times 10^{-16}~\mathrm{erg~s^{-1}~cm^{-2}~\text{\AA}^{-1}}$ covered by the bandwidth of the KP1468 filter \citep{Hanuschik03}.
Due to the fact that the intensities of sky emission lines vary considerably with time \citep{Roach73}, we generally can observe part of these lines in one exposure.
For an example shown in Fig.~\ref{fig:sky_line_shift}, we identify three sky lines ($\lambda = 6533.05, 6553.63, 6562.76$~\text{\AA}) from our spectra produced by each long slit and measure their peak wavelengths by fitting Gaussian functions.
We then calculate the shift between the wavelengths obtained from the best-fit and the sky line template for each long slit, and use this shift to further correct for the bias in the above wavelength calibration using the lamp lines. 

\end{CJK*}
\end{document}